\documentclass[12pt,preprint]{aastex}

\newcommand{\che} {\log\ ({\rm C/He})}
\newcommand{\ch} {\log\ ({\rm C/H})}

\newcommand{\msun} {$M_{\odot}$}

\newcommand{\Te} {T_{\rm eff}}
\newcommand{\logg} {\log g}

\begin{document}

\title{Hot DQ White Dwarfs: Something Different}

\author{P. Dufour\altaffilmark{1},
G. Fontaine\altaffilmark{2},
James Liebert\altaffilmark{1},
G. D. Schmidt\altaffilmark{1},
N. Behara\altaffilmark{3}}

\altaffiltext{1}{Steward Observatory, University of Arizona, 933 North Cherry Avenue, Tucson, AZ 85721; dufourpa@as.arizona.edu, liebert@as.arizona.edu, schmidt@as.arizona.edu}
\altaffiltext{2}{D\'{e}partement de Physique, Universit\'{e} 
de Montr\'{e}al, C.P. 6128, Succ. Centre-Ville, Montr\'{e}al, Qu\'{e}bec, 
Canada H3C 3J7; fontaine@astro.umontreal.ca}
\altaffiltext{3}{CIFIST, GEPI, Observatoire Paris-Meudon, 92195, France; natalie.behara@obspm.fr}

\begin{abstract}

We present a detailed analysis of all the known Hot DQ white dwarfs in
the Fourth Data Release of the Sloan Digital Sky Survey (SDSS)
recently found to have carbon dominated atmospheres. Our spectroscopic
and photometric analysis reveals that these objects all have effective
temperatures between $\sim$ 18,000 and 24,000 K. The surface
composition is found to be completely dominated by carbon, as revealed
by the absence of H$\beta$ and He~\textsc{i} $\lambda$4471 lines (or
determination of trace amount in a few cases). We find that the
surface gravity of all objects but one seems to be ''normal'' and
around $\logg$ = 8.0 while one is likely near $\logg$ = 9.0. The
presence of a weak magnetic field is directly detected by
spectropolarimetry in one object and is suspected in two others.  We
propose that these strange stars could be cooled down versions of the
weird PG1159 star H1504+65 and form a new family of hydrogen and
helium deficient objects following the post-AGB phase. Finally, we
present the results of full nonadiabatic calculations dedicated
specifically to each of the Hot DQ that show that only SDSS
J142625.70+575218.4 is expected to exhibit luminosity variations. This
result is in excellent agreement with recent observations by
Montgomery et al.  who find that J142625.70+575218.4 is the only
pulsator among 6 Hot DQ white dwarfs surveyed in February 2008.

\end{abstract}


\section{INTRODUCTION}

White dwarfs represent the final stage of stellar evolution for the
vast majority of stars that have exhausted the nuclear fuel available
in their core (this is the fate of $\sim 97 \%$ of stars in our
Galaxy). Standard stellar evolution theory predicts that a typical
white dwarf is composed of a core that encompasses more than 99 $\%$ of
the mass of the star, surrounded by a thin envelope of helium (and
hydrogen) that has survived the nuclear burning and mass loss
phase. The core, which is essentially the result of the fusion of
light elements, has a composition that depends on the initial mass of
the star. For very low mass stars that could not ignite helium, it
is composed of helium. Stars at intermediate masses end up with
cores composed of carbon and oxygen while the most massive stars
produce oxygen-neon-magnesium (ONeMg) cores. The initial mass function
and star formation history of the Galaxy are such that the majority of
white dwarfs we observe in the solar neighborhood today have a core
made of carbon and oxygen  \cite[the most massive white dwarfs are
intrinsically rare while low mass stars producing helium core white
dwarfs have nuclear lifetimes greater than the age of the disk of our
Galaxy and are thus possible only from binary evolution or if the mass
loss in the red giant phase is extremely large,][]{kalirai07}.

Direct observation of the core is unfortunately not possible since it
is surrounded by a thin and opaque layer of helium \citep[and hydrogen
for $\sim$ 80 $\%$, see][]{iben84,koester86,dantona87} that was left over
in the the previous stage of stellar evolution. White dwarfs have thus
traditionally been separated into two distinct families: those with a
hydrogen rich surface composition, and those with a helium rich
surface composition. The former are represented by the very well known
DA white dwarfs while the latter family show more diversity in their
spectral type. Indeed, although they all have a helium rich surface
composition, they have been subdivided into several spectral types
which reflect essentially their effective temperatures. The hottest ones 
($\Te > \sim$ 40,000 K) that show He~\textsc{ii} lines are classified
DO. At intermediate temperature ($\sim 12,000 - 40,000$ K), those
showing exclusively He~\textsc{i} lines are classified as DB. Stars too 
cool for the He~I atoms to be sufficiently excited and show no
features in their optical spectra are labeled DC. Finally, the cool
helium dominated stars with traces of carbon (either molecular or
atomic) in their optical spectra that are found between $\sim$ 4000 K
and 13,000 K are classified as DQ white dwarfs.

Carbon is also observed in the spectra of the hot PG1159 stars,
objects just entering or about to enter the white dwarf cooling
phase. Their atmospheres show a mixture of helium, carbon and oxygen
and little or no hydrogen \citep[see][and references 
therein]{werner06}. The most likely explanation for this unusual
composition is that these stars have experienced the so-called ''born
again'' scenario, i.e. a very late thermal pulse that has brought back
the white dwarf onto the post-AGB phase for a second time early in its
cooling phase \citep[see][and references therein]{herwig99}. As a
result of this rather violent event, the star re-enters the white
dwarf cooling phase but this time with a surface composition that is
devoid of hydrogen. This process also mixes the remaining helium with
elements from the envelope (mostly carbon and oxygen), producing the
curious surface composition observed in PG1159.  Gravitational
diffusion will eventually separate the helium from the heavier
elements and PG1159 stars will thus turn into helium rich DO, then DB and
finally DQ white dwarfs as the cooling continues.

The presence of carbon in cool DQ white dwarfs is now well explained
by a model in which carbon diffusing upward from the core is brought
to the photosphere by the deep helium convection zone
\citep{pelletier86}. Detailed analysis of a large sample of 56 DQ
stars by \citet{dufour05} showed that $\che$ varies from about $-$7 at
$\Te \sim 6000$ K to $-$3 at $\Te \sim 11,000$ K. The maximum
contamination by carbon is predicted to be at $\Te \sim$
12,000 K, i.e when the convection zone reaches its maximum. The only
two stars close to this maximum that have been analyzed, G35-26
\citep{thejll90} and G227-5 \citep{wegner85}, have respectively $\Te$
of 12,500 $\pm$ 1,500 K and 12,500 $\pm$ 500 K and $\che$= $-$2 and
$-$2.5. Parallax measurements (Dahn, private communication) have since
shown that these two stars are massive white dwarfs ($\sim 1.1$
\msun). At these temperatures and abundances, the spectra show
primarily C~\textsc{i} lines and a hint of the He~\textsc{i}
$\lambda4471$ line that is about to fade below visibility as the stars
become too cool. While the cool range of the DQ distribution has been
extensively analyzed thanks to the large sample from the SDSS
\citep{dufour05,koester06}, our knowledge of carbon abundances for
hotter stars relies only on the two stars mentioned above and a few
analyses of DB stars that show carbon only in the ultraviolet
\citep{provencal00,dufour02,petitclerc05,desharnais08}.

The situation can now be improved dramatically thanks again to the
discovery of several new hot DQ white dwarfs in SDSS
\citep{liebert03}. Most of these stars have spectra similar to those 
of G35-26 and G227-5 and show essentially only C~\textsc{i} lines (as
opposed to molecular bands of C$_2$ for cooler stars). Surprisingly,
some stars showed spectra that were dominated by C~\textsc{ii} lines. No 
detailed analysis could be performed at the time due to the lack of
proper models for this type of stars. Estimates of effective
temperatures based on pure helium models placed these stars well into the
DB temperature range, although the exact atmospheric parameters were
uncertain since analysis with pure helium models tends to overestimate
the effective temperature \citep{provencal02, dufour05}. It was presumed 
that they were simply hotter versions of DQ stars like G35-26 and
G227-5. \citet{liebert03} hypothesized that the dredge up of carbon for
these special stars had happened earlier than usual because they were
more massive and had a thinner outer helium layer, as the theoretical
calculation of \citet{kawai88} show. It was believed that the increased
continuum opacity resulting from the presence of the carbon was keeping
us from detecting the strong helium lines that we usually see in DB
spectra. According to the \citet{liebert03} scenario, the most massive
DB stars simply dredged up carbon at hotter effective temperature, which
made them to appear as hot DQ stars. The \citet{beauchamp95} and
\citet{voss07} analyses of large samples of DB white dwarfs show that
the DB mass distribution lacks the high mass component that is observed
for DA white dwarfs. The scenario above was thus proposed as a natural
explanation to the fact that the DB mass distribution morphology is
fundamentally different from that of DA white dwarfs.

However, when we proceeded with the calculation of the appropriate
models, we found that if helium was the dominant atmospheric
constituent, it remained spectroscopically observable even when
opacity due to a large quantity of carbon was accounted for. As a
consequence, it seems that these stars have an atmosphere that is
dominated by carbon with little or no trace of other elements. We
could thus be directly witnessing the bare stellar CO core,
or a carbon envelope around an ONeMg core, giving us a unique
opportunity to test theories of stellar evolution. Note that standard
dredge-up theory cannot explain this phenomenon since at the effective
temperatures where these stars are found, the helium convection zone
doesn't reach deep enough to bring a significant amount of carbon to
the surface. Another explanation is thus required. According to the
standard spectral classification defined in \citet{mccooksion99},
white dwarfs that show carbon features, either atomic or molecular in
any part of the electromagnetic spectrum are referred to as DQ. In order
to avoid confusion with the cooler DQs, we will thus simply refer to
stars that show mainly ionized carbon features as ''Hot DQs'',
even though they are truly something different. The surprising
discovery of these objects has been published recently in the form of
a short letter \citep{dufourNat}. We now wish to present in
more detail an up to date analysis of all the carbon dominated
atmosphere white dwarfs that we have uncovered thus far as well as an in
depth discussion of the possible origin and and evolution of such
objects.

In \S~\ref{observation}, we describe the observations. Our theoretical
framework including our model atmosphere and synthetic spectrum
calculations are presented in \S~\ref{theoretical}. The detailed
analysis follows in \S~\ref{analysis}, and the results are
interpreted and discussed in \S~\ref{results}. Our conclusions are
summarized in \S~\ref{conclusion}.

\section{OBSERVATIONS}\label{observation}

Our sample consists of white dwarf stars spectroscopically identified as
''Hot DQ'' in the SDSS Fourth Data Release white dwarf catalog of
\citet{eisenstein06}. Note that a fraction of the Hot DQs in this catalog
are actually DQ white dwarfs in the 12,000-15,000 K temperature range
that have helium-rich surface compositions ($\che \sim -2$ to $-3$),
and we do not include them in the present analysis (they will be
analyzed in detail in Dufour et al., in preparation). Details concerning
the observations can be found in \citet{eisenstein06} and reference
therein. Our final sample consists of 9 Hot DQ stars with SDSS spectra
covering the 3800-9200 \AA\ region at a resolution of $\sim$ 3 \AA\
FWHM. Also available are SDSS photometric observations on the $ugriz$
system \citep{fukugita96, hogg01,smith02,ivezic04}. The spectra of a few
of these stars have been previously shown in \citet{liebert03}.

\section{MODEL ATMOSPHERE AND SYNTHETIC SPECTRUM CALCULATIONS}\label{theoretical}

The LTE model atmosphere code used for this analysis is similar to
that described in \citet{dufour05,dufour07} and references therein for
the study of cool DQ and DZ white dwarfs. It is based on a modified
version of the code described at length in \citet{BSW95}, which is
appropriate for pure hydrogen and pure helium atmospheric
compositions, as well as mixed hydrogen and helium compositions, while
energy transport by convection is treated within the mixing-length
theory. One important modification is that metals and molecules are
now included in the equation-of-state and opacity calculations
\citep[see][for details]{dufourphd}. The principal modification in the
version used in this paper concerns the treatment of the continuum
opacity of heavy elements. In our earlier version of the code, the
bound-free and free-free opacities were taken from the \citet{peach70}
opacity tables. In the case of cool helium-rich white dwarf
atmospheres (DQ, DZ and DC spectral type), the exact treatment of the
metallic continuum opacity is not really important since the dominant
source of opacity is the He$^-$ free-free absorption. However, for
hotter carbon-rich stars where the far-ultraviolet flux is far from
negligible, it becomes important to include the best possible sources
of opacity. Small variation in the absorption at short wavelengths can
significantly modify the flux redistribution in these atmospheres as
was shown by \citet{behara05}. Our models now include the latest
photoionization cross-sections from the \citet{OP95,OP97} for
C\textsc{i}-\textsc{iv} and O\textsc{i}-\textsc{iv} as well as the
corresponding free-free absorption as described in detail in
\citet{behara06}. Also, our earlier version of the code was dedicated
to the analysis of cool white dwarfs in a regime where helium was
spectroscopically invisible, and as a consequence, no effort was made
to keep it up to date with the best He~\textsc{i} line opacities
\citep[those of][]{beauchamp97}. The present version has thus been
updated to include these He~\textsc{i} line profiles. Since the
charged carbon particles impose an electric micro-field at the
location of the radiating atom, we also include a correction to the
critical field strength and the plasma correlation parameter in the
occupation probability formalism \citep{HM88} as described in
\citet{werner99}.

The strongest C and O lines are included explicitly in both the model
and synthetic spectrum calculations. These lines are selected by
taking all lines contributing more than one tenth of the continuum
opacity at each optical depth. A test model calculated by
including all the lines from our list was not found to yield any
detectable difference on the emergent spectrum. The line absorption
coefficient is calculated using a Voigt profile for every line at
every depth point. Central wavelengths of the transitions, $gf$
values, energy levels, and damping constants are extracted from the
GFALL line list of R.~L.~Kurucz\footnote{see
http://kurucz.harvard.edu/LINELISTS.html}.

Two grids with different C/He and C/H abundance have been calculated.
Our first grid covers a range from $\Te$ = 16,000 to 30,000 K in steps
of 2000 K, from $\logg$ = 7.5 to 9.0 in steps of 0.5 dex, and from
$\che$ = +3.0 to 0.0 in steps of 1.0 dex. Our second grid covers the
same parameter space except that it is $\ch$ varying from +3.0 to
0.0 in steps of 1.0 dex. Finally, additional models with various
abundances of oxygen have been calculated to explore the sensitivity of
our results to this unknown parameter. Illustrative spectra from our
grid are displayed in Figure \ref{fg:f1} for various values of the effective
temperature, gravity, hydrogen and helium abundances.

\section{DETAILED ANALYSIS}\label{analysis}

\subsection{Fitting Technique}

The effective temperatures of the stars can be determined from either the
photometric energy distributions or from fits to the carbon lines. In
both cases, the fitting procedure relies on the nonlinear
least-squares method of Levenberg-Marquardt \citep{pressetal92}.

In the first case, we fit the $ugriz$ colors with both $\Te$ and the
solid angle $\pi(R/D)^2$, which relates the flux at the surface of
the star to that received at Earth (R is the radius of the star and D
its distance from Earth), as free parameters. This is done for a fixed
value of $\logg$ and $\che$ or $\ch$ (we will explore the effect of
the former below while the latter are constrained/determined from the
absence/presence of spectroscopic features). Corrections of the
photometric measurements to account for the extinction from the
interstellar medium are obtained from the reddening maps of
\citet{schlegel98}. We experimented with various fractions of the
absorption in the line of sight to assess the effect of this unknown
parameter on our atmospheric parameter determinations (see
below). Finally, we compare the observed spectra with the
synthetic spectra interpolated at the solution
obtained from the photometric fit.

In the second case, we do the inverse. That is we first fit the carbon
lines from the spectroscopic data to obtain the effective temperature
(with the solid angle and the slope, to account for the unknown amounts 
of reddening and spectrophotometric errors, left as free parameters)
and then fit the photometric data with $\Te$ fixed to that value to
obtain the solid angle (this gives the photometric distance, see
Table~1) and assess the quality of the solution. Again, this is done
for various fixed values of $\logg$ and $\che$ or $\ch$.

\subsection{Hydrogen and Helium Abundances}

With the possible exception of one star, none of the Hot DQ stars
analyzed in this paper show the presence of helium (He~\textsc{i}$
\lambda$4471). Therefore, only upper limits can be set from the
spectroscopic observations. These limits depend sensitively on the
range of effective temperature considered. We roughly estimate these
limits by inspecting our grid of synthetic spectra. We estimate,
for the typical signal-to-noise ratio of the SDSS observations, that at
$\logg$ = 8, the threshold for the detection of helium as a function
of effective temperature is at $\che \sim$ 0.7 for $\Te$ = 18,000 K,
$\sim$ 1.6 for $\Te$ = 22,000 K and $\sim$ 1.8 for $\Te$ = 26,000
K. At higher gravity, the $\lambda$4471 line becomes very broad and
much more helium can be present without being observable. For example,
at $\logg$ = 9, the limits are $\che \sim$ 0.0 for $\Te$ = 18,000 K
while at $\Te$ = 22,000 K the limit is $\sim$ 1.0.

The limits on the amount of hydrogen are more difficult to assess, 
since the H$\alpha$ line is contaminated by strong carbon
features. The H$\beta$ line is thus used to set the limit of
visibility of hydrogen since it is free of contamination for most of
the range of effective temperature we are exploring. We find that
H$\beta$ is spectroscopically observable for $\ch \sim$ 2.7 for $\Te$
below $\sim$ 23,000 K. At higher effective temperature, a carbon line
starts to become strong at the position of H$\beta$ and no limit can
be obtained with complete certainty. 

We note that the influence of small, spectroscopically invisible 
amounts of hydrogen or helium do not affect significantly the
effective temperature determination (except for the coolest star,
see below) and that fits without hydrogen or helium are practically
identical to those with H and He set at the limits of visibility.

\subsection{Atmospheric Parameter Determinations}

\noindent {\it SDSS J000555.90$-$100213.3 :} The spectrum of this star
shows an obvious sign of magnetic line splitting. The $\sim$4270 \AA\
line seems to be split into three components approximatively 25 \AA\
apart, corresponding to a mean surface field strength of
$B_S\approx1.47$~MG \citep[see eq. 3 in][where an effective Land\'e
factor equal to unity is assumed]{dufour06}. Circular
spectropolarimetry obtained with the spectropolarimeter SPOL
\citep{schmidt92} on the 2.3 m Bok reflector at Kitt Peak confirms the
presence of a magnetic field on {\it SDSS J000555.90$-$100213.3}.
These data were acquired on 2007 Oct. 17 and Dec. 14 using a single
order of a low resolution grating ($\Delta\lambda\sim17$~\AA) to cover
the region 4000$-$8000~\AA.  Polarimetric sequences each totaling
4800~s were obtained both nights, and because the results were
indistinguishable within the statistical noise, the average is
displayed in Figure \ref{fg:f2}.  Clear polarization reversals are
seen around the prominent C~\textsc{ii} lines at
$\lambda\lambda$4267,4370, near $\lambda$4860, and, with somewhat less
significance, at the broad, shallow $\lambda$6578,6583 C~\textsc{ii}
feature.  The $\lambda$4860 polarimetric feature is real, as it is
present in both data sets, despite the fact that a strong absorption
line is not observed here.  However, inspection of the flux spectra
from the SDSS as well as from our data reveal that the region is
clearly affected by a number of weaker lines.  The longitudinal
magnetic field is difficult to estimate from the data because the
Zeeman splitting is actually somewhat less than the separation between
the two components that comprise the $\lambda$4300 feature, and the
$\lambda$6580 line is both shallow and noisy.  However, simulations of
the blue feature using reasonable intrinsic line profiles suggest that
a mean longitudinal value $B_e\sim400-800$~kG is appropriate for the
data.  This is consistent with the estimate of mean surface field
$B_S\approx1.47$~MG based on the observed splitting of the line core
into three Zeeman components. The presence of the magnetic field seems
to have destroyed all the carbon features that are strongly seen in
most other objects of our sample. As a consequence, we cannot estimate
the effective temperature by fitting the lines and thus use the
photometric data. For the same reason, we cannot constrain or estimate
the surface gravity for this star. We thus assume $\logg$ = 8.0. Our
fit of the energy distribution yields $\Te$ = 17,400 K if no reddening
correction is applied and $\Te$ = 19,420 K if we assume the full
correction from the \citet{schlegel98} maps. The corresponding fit to
the solid angle places this star at a distance of $\sim$ 250 pc. A
fair fraction of the extinction on that line of sight should thus be
applied so we believe that the hottest solution is probably more
realistic. This is this solution that we show in Figure \ref{fg:f3}
and Table~1.

\noindent {\it SDSS J010647.92+151327.8 :} The extinction in the line of
sight to this star is particularly high and as a result, the 
photometric fits with and without correction for the extinction are very
different ($\Te$ of respectively 17,900 and 22,900 K). The synthetic
spectra interpolated to the photometric solution without correction for
the extinction is clearly at odds with the observed spectra while the
hotter solution looks better even though it suffers from obvious
flaws. First, the 4270 line is predicted to be too strong and the
features near 4100$-$4200 \AA\ are also poorly reproduced. There might
also be a small amount of hydrogen as suggested by the feature near
H$\beta$.  If we believe the small absorption near $\lambda$4471, this
object might be the only one in our sample showing a trace of
helium. We can approximately reproduce the $\lambda$4471 feature if we
allow $\che \sim$ 1.0. Fits with this abundance yield $\Te$ = 23,430 K,
making this star the hottest of our sample. It is possible that the
lines are also affected by the presence of a weak magnetic field, but
polarization measurements are needed to confirm this hypothesis. We thus
do not attempt to fit the lines individually and only give our $\logg$ =
8.0, $\che$ = 1.0 photometric fit with full correction for extinction in Figure
\ref{fg:f3} and Table~1.

\noindent {\it SDSS J023637.42$-$073429.5 :} Our $\logg$ = 8.0
photometric fits with and without correction for the extinction give
respectively $\Te$ = 22,960 and 20,780 K. Both synthetic spectra
interpolated at these solutions compare reasonably well with the
observations considering the signal-to-noise ratio of the spectrum. It
is reassuring that if we use only the spectra in our minimization
technique, we get an effective temperature that is intermediate at
$\Te$ = 21,100 K. As for the surface gravity, a value as high as
$\logg$ = 9.0 is clearly ruled out, since the resulting line profiles
would be far too broad to be compatible with the observation. However,
a value of $\logg$ = 8.5 cannot by dismissed but a higher
signal-to-noise ratio observation will be needed to better constrain
this parameter.

\noindent {\it SDSS J115305.54+005646.2 :} Our $\logg$ = 8.0
photometric fits with and without correction for the extinction give
respectively $\Te$ = 21,400 and 20,160 K. Inspecting the resulting
synthetic spectra, we find that the hotter solution looks is a
slightly better match. The solution by fitting the lines confirms
that result, since it yields $\Te$ = 21,650 K. Although the overall
appearance of the fit looks very good (see Figure \ref{fg:f3}),
several discrepancies catch our attention. In particular, we notice
that the strengths of the four weaker lines that follow the strong
$\lambda$4270 are not well reproduced by our model. The first is
barely seen in our synthetic spectra, while the second is too strong.
The third and fourth have about the right strength, although the fit
is far from perfect.  We note also that the strong feature near $\sim$
5130 \AA\ is poorly reproduced. Finally, the line near $\sim$ 4960
\AA\ is not predicted at all by our models for this effective
temperature (it starts to appear only at about 26,000 K, see Figure
\ref{fg:f1}). In fact, if we look carefully at Figure \ref{fg:f1}, we
notice that all the above mentioned lines have the right strength only
at a much higher effective temperature than what we have determined
here. It seems as if that these few lines, and only these few ones,
required a higher effective temperature. We will discuss in more
detail in the next section possible explanations for these
discrepancies. As for surface gravity, as it is the case for all the
objects in our sample, the quality of the data at hand (as well as the
problem with the strengths of some lines noted above) does not allow
us to precisely determine this parameter. However, the value of
$\logg$ is probably not too far from 8.0 since for fits with $\logg$
fixed at 8.5, we already find that the lines are broader than
observed.

\noindent {\it SDSS J133710.19$-$002643.7 :} This star is very
similar to {\it SDSS J023637.42$-$073429.5}, except for the clear
presence of hydrogen as revealed by the H$\beta$ line. Our $\logg$ =
8.0 photometric fits with and without correction for the extinction
give respectively $\Te$ =  22,550 and 20,770 K. Both synthetic spectra
interpolated at these solutions compare reasonably well with the
observation, considering the signal-to-noise ratio of the
spectrum. Again, it is reassuring that if we use only the spectra in
our minimization technique, we get an effective temperature that is
intermediate at $\Te$ = 21,670 K. The value $\ch$ we get for that
temperature is 1.67. The surface gravity seems to be near $\logg$ =
8.0 and perhaps even less. A surface gravity as high as $\logg$ = 8.5
is completely ruled out, since that would produce lines that are too
broad to be compatible with the observation. We note that again, we
observe some discrepancies in the line strength of the four lines
long-ward of the $\sim$ 4270 line and the line near $\sim$ 4960.

\noindent {\it SDSS J142625.70+575218.4 :} The spectrum of this star
doesn't show as many carbon lines as we see in most other
objects. There is a strong feature near $\sim$ 4270 \AA\ and hints of
absorption for a few more lines. Our $\logg$ = 8 fit indicates that
this star is one of the coolest of the sample with an effective
temperature of about 17,000-17,500 K. However, optical synthetic
spectra with such parameters are predicted to show many strong carbon
features, in particular the group of lines long-ward of $\sim$ 4270
\AA\, that are not observed in the SDSS spectrum. Inspection of Figure
\ref{fg:f1} indicates that synthetic spectra having similar
characteristics to {\it SDSS J142625.70+575218.4} exist for a higher
$\Te$ and $\logg$. We believe we can thus reject a low surface gravity
solution and that $\logg$ near 9.0 is more likely. If we fix $\logg$
at 9.0, we get a good fit of the energy distribution at $\Te$ = 17,800
K but the lines long-ward of $\sim$ 4270 \AA\ still remain too
strong. If we use only the spectrum in our minimization procedure, we
achieve a much better fit at $\Te$ = 20,500 but then the u, r and z
band fluxes are poorly reproduced. At high gravity and low effective
temperature, much more helium can be hidden without being
spectroscopically observable. In this particular case, the presence of
helium does affect slightly our atmospheric parameter
determination. If we assume $\logg$ = 9.0 and $\che$ = 0.0, we obtain
a very acceptable fit for the spectrum at $\Te$ = 19,830 K and only
the u band is slightly over predicted. This solution (this is the one
we present in Figure \ref{fg:f3} and Table~1) also predicts a tiny
depression from the He~\textsc{i} $\lambda$4471 line that is
compatible with the observations (the spectrum seems to show a
possible glitch at the right position but better signal-to-noise ratio
observations are required to confirm the presence of helium). It is
also possible that the surface gravity exceeds 9.0 and that a better
solution exists outside the parameter space of our grid. Finally, we
notice that our solution does not predict correctly the strength of
the $\sim$4620 \AA\ line, suggesting perhaps that the effective
temperature might actually be a little higher.

\noindent {\it SDSS J161531.71+454322.4 :} This star is one of the few
that shows the presence of a H$\beta$ line. Our photometric fits with and
without correction for extinction give an effective temperature of,
respectively, 18,300 and 17,600 K. However, these solutions predict an
optical spectrum that is clearly at odds with the observation. The
strengths of the observed lines seem to indicate that the effective
temperature is much higher. If we use only the spectrum in our
minimization procedure, we get a much better fit with $\Te$ = 20,940 K
and $\ch$ = 1.74. The price to pay with this solution is that the u band
flux is predicted to be much stronger than observed. Higher surface
gravity solutions are ruled out as even for $\logg$ = 8.5, the predicted
line profiles are too broad.

\noindent {\it SDSS J220029.09$-$074121.5 :} Our $\logg$ = 8.0
photometric fits with and without correction for the extinction give 
respectively $\Te$ = 21,200 and 18,500 K. From the inspection of the
predicted optical spectra at these solutions, we can easily reject the
coolest of the two, as the line strengths are then poorly 
reproduced. If we fit only the lines, we get 21,240 K and this is
thus the solution that we present in Figure \ref{fg:f3} and
Table~1. We note that the observed $\sim$ 4270 \AA\ line is much
broader than predicted, indicating perhaps that the surface gravity is
higher than $\logg$ = 8.0. However, if we try to fit this star with a
higher gravity, we get a better fit to the $\sim$ 4270 \AA\ line but
then all the other lines are too broad. One attractive possibility is
that the line profiles for this star are affected by the presence of a
weak magnetic field but the lines are not separated enough to be
individually identified. Polarimetric measurement should easily
confirm the validity of this hypothesis. 

\noindent {\it SDSS J234843.30$-$094245.2 :} The $\logg$ = 8.0
photometric solutions with and without correction for extinction give
respectively $\Te$ = 17,600 and 16,000 K. At these temperatures, the
predicted optical spectra are clearly at odds with the observations. A
much better fit at $\Te$ = 21,550 K can be obtained by using only the
line profiles in our minimization procedure. However, the corresponding
energy distribution with this solution is poorly reproduced. We have no
explanation for such a discrepancy. A higher surface gravity is again
ruled out, as the predicted line profiles would be too broad. We note
again that we observe some discrepancies in the line strength of the
four lines long-ward of the $\sim$ 4270 line, the line near $\sim$ 4960
and another near $\sim$ 5260 \AA\ .

\subsection{Source of Uncertainties}

\subsubsection{Flux in the ultraviolet}

At the effective temperatures where the carbon dominated atmosphere
white dwarfs are found, a large fraction of the light is emitted in
the UV portion of the electromagnetic spectrum. Figure \ref{fg:f4}
shows two synthetic spectra at $\Te$ = 22,000 K for pure helium (red)
and a pure carbon atmosphere (blue). The most striking characteristic
is the strength and number of all the carbon features in the UV part of
the spectra in the pure carbon model while the DB model is practically
featureless in the same region. These absorptions in the UV have a
huge impact on the energy distribution. This can easily be appreciated
by remembering that the total energy emitted, $\int H_{\nu} d\nu$, is
by definition, for a given $\Te$, the same for both models (= $\sigma
\Te ^4/4\pi $). All the flux below $\sim$ 1500 \AA\ that is absorbed
in the pure carbon model, compared to the DB model, is redistributed
at longer wavelengths. It is thus important to include as accurately
as possible all sources of opacity in the model calculation, as errors
in our treatment of the absorption in the UV might have significant
impact on the thermodynamic stratification and the overall energy
distribution. These errors can be of multiple origins. The line list we
use can be incomplete, values of log gf for some lines might be wrong
by unknown factors, unexpected absorption from other elements might be
present, or the broadening theory we use might be incorrect (see below
for more on that issue). It is thus quite possible that the above
mentioned discrepancies observed for many lines in the optical might
be related to such errors (alternatively, the few lines that have log gf
errors in our list might just stand out).

Therefore, until UV observations are obtained to assess the quality of
our modeling of that part of the spectrum, our atmospheric parameter
determinations remain uncertain. It is probable that, as we learn
more on these various aspects, future generation of models will have
different thermodynamic stratification which could yield slightly
different atmospheric parameters. It is difficult to quantify this for
the moment, since no UV observations are yet available, but given the
inconsistencies enumerated above (i.e. the few lines that are poorly
reproduced), it is possible that our effective temperature
determinations could perhaps be wrong by as much as 2000 K, a larger
value than what is reflected in the formal statistical uncertainties
presented in Table~1.

On the other hand, due to the richness of the UV part of the spectra,
these stars give us a unique opportunity to test the accuracy of the
theoretical atomic data, opacity sources and broadening theory for the
carbon atom since we are observing carbon in physical conditions and
abundances never seen previously. Detailed comparison of state of the
art models with future UV observations should allow us to better
assess the quality of our modeling and reduce significantly the
uncertainties on our derived atmospheric parameters.

\subsubsection{Influence of Oxygen}

Since oxygen is also expected to be present in large proportion, along
with carbon, in the cores of white dwarf stars, we discuss in this
section how it could influence our results if present in the
atmosphere. In figure \ref{fg:f5}, we show synthetic spectra of $\logg$
= 8, $\Te$ = 22,000 K atmosphere models with a pure surface composition
of hydrogen, helium, carbon and oxygen. We first notice that for a pure
oxygen model, the optical spectrum is not as rich in spectral features
as those made with lighter elements. With the exception of a few strong
lines, most of these features would not even be identifiable in a
spectrum as noisy as our SDSS Hot DQ stars. Inspection the 9 Hot DQ
spectra did not reveal any evidence of an oxygen feature at the position
of the strong oxygen lines seen in Figure \ref{fg:f5}. Models assuming a
surface composition of 50\% C and 50\% O are also very similar to those
of 100\% C, except that the strongest oxygen lines can be noticed and
that the carbon lines appear a little bit broader. However, since we do
not observe these oxygen lines in any of our spectra, the oxygen
abundance has to be at least close to an order of magnitude inferior to
that of carbon. A more precise limit will require the analysis of higher
signal-to-noise ratio observations, but since the C/O ratios are certainly
much higher than one, this practically guarantees that the atmospheric
parameters derived in this study should not significantly be altered by
the presence of an unknown but small amount of oxygen.

\subsubsection{Stark Broadening}

The dominant source of broadening in the conditions encountered in
these type of stars is Stark broadening. In this first generation
of carbon atmosphere models, we made use of the so-called ''scaled
classical approximation'' \citep{kurucz81,hubeny88}. This
approximation assumes $\Gamma_S/N_e$ to be a constant. How realistic
this approximation is under the atmospheric conditions met here
remains to be tested. Uncertainties associated with the precision of
the treatment of the line broadening of C~\textsc{ii} could thus prove
to be a large source of error in our model calculations and surface
gravity estimations. As discussed above, errors in the absorption
coefficient can have a particularly strong effect in the UV part of
the spectrum, not to mention the line profiles in the optical. We thus
believe we should remain very cautious in our interpretation of the
limits/determinations of the surface gravity as derived from the line
profiles. We are currently working on incorporating the best Stark
broadening theory available for C~\textsc{ii} lines in our model
calculations and the results will be presented in due time.

\section{DISCUSSION}\label{results}

\subsection{Spectral Evolution Scenario}

In this section, we present what we believe is a most likely scenario
for the formation and evolution of carbon atmosphere white dwarfs. The
main line of our argument has been briefly discussed in
\citet{dufourNat} and \citet{dufourhydef}.

The origin of hydrogen deficient stars is generally believed to be
explained by a late helium-shell flash where a post-AGB star (or white
dwarf) re-ignites helium-shell burning. This brings the star back onto the
AGB phase (the ''born again'' scenario) and the associated envelope
mixing and mass loss eliminate the remaining hydrogen. As a result, the
surface composition of these objects will be a mix of helium, carbon and
oxygen. PG1159 and non-DA (helium-rich) stars are probably
descendants of objects that have experienced this late flash. It is thus
certainly not too far-fetched to imagine that a particularly violent
late thermal pulse could also explain the origin of stars that are both
hydrogen {\it and} helium deficient. The hottest PG1159 star H1504+65
($\Te \sim$ 175,000 K) might actually owe its origin to such a
scenario. Its surface composition is nothing short of amazing with a
mass fraction of C$\sim 48\%$, O $\sim 48\%$, Mg $\sim 2\%$ and Ne $\sim
2\%$ and until the recent discovery of the Hot DQ white dwarfs, H1504+65
was the only known object to not show traces of either hydrogen or
helium \citep{werner91,werner99,werner04}. It is thus very tempting to
draw an evolutionary link between H1504+65 and the cooler carbon
dominated atmosphere white dwarfs.

As a star like H1504+65 cools, it is expected that
gravitational diffusion will separate carbon from oxygen so that the
atmosphere should rapidly appear carbon dominated. One flaw to this
simple scenario is that we would expect to find carbon dominated
atmospheres all along the cooling sequence, not only in a narrow strip
centered around $\sim$ 20,000 K. However, a careful search through the
SDSS archive didn't reveal any new carbon dominated white dwarfs with
an effective temperature higher than those presented in this paper (we
looked at all the DR6 spectra with $u-g < -0.2$ and $g-r <
-0.3$). Thus, even though our sample is small, it is certainly
significant that {\it all} objects are found in a narrow effective
temperature strip and {\it none} at higher or lower temperature. Any
proposed scenario must be able to explain this extraordinary fact.

On the other hand, we do not believe that this necessarily mean that
there is no possible relationship between stars like H1504+65 and the
Hot DQs. Even though helium is not detected in the atmosphere of
H1504+65, it doesn't mean that none is present since this star is so hot
that a small amount of helium could be present but completely ionized,
preventing its detection from spectral analysis \citep{nousek86}. Hence,
the small residual amount of helium possibly existing in the envelope of
H1504+65 should eventually diffuse upward to form a thin layer above a
carbon-enriched and oxygen-depleted mantle. Since the total mass of an
atmosphere is tiny ($ \sim10^{-14}-10^{-15}$ \msun), there should
ultimately be enough accumulated helium to form a full atmosphere and
the descendant of H1504+65-like stars should then ''disguise''
themselves as He-atmosphere white dwarfs following the PG1159
evolutionary phase. We believe that these helium-rich stars could then
cool normally as DO/DB stars until a convection zone develops in the
carbon-enriched mantle due to the recombination of that element. At that
point in time, the subphotospheric carbon convection zone becomes active
enough to be able to dilute from below the very thin overlying radiative
He layer and the star undergoes a dramatic spectral change, transforming
itself from a He-dominated atmosphere white dwarf to a carbon-dominated
atmosphere star.  Since the mass in the carbon convection zone is orders
of magnitude larger than the mass of the He layer, helium would become
spectroscopically invisible. The exact value at which this dilution take
place is not known due to a current lack of proper modelling, but it is
conjectured that it should be around 24,000 K.  SDSS
J010647.92+151327.8, the hottest and the only star clearly showing
helium in our sample could perhaps be a rare star caught in the act of
converting from a DB to a Hot DQ.

In addition, the carbon dominated atmosphere phase in the evolution of
such a star must be short-lived as no carbon dominated cases are found for
effective temperatures lower than $\sim$ 18,000 K. Such stars, if they
existed, would not only be easily recognizable from their spectra (the
atomic or molecular features would be extremely strong, see
Fig. \ref{fg:f6}), but should also be more numerous since the cooling
times are much longer as we approach the low end of the cooling sequence
(in other word, stars tend to accumulate at low effective temperature
while they pass rapidly through the hotter phase). The fact that we
haven't uncovered them probably means that helium must ultimately
reappear at the surface in enough quantity to form again a helium-rich
atmosphere (high signal-to-noise ratio observation revealing the
presence of helium in one of the coolest stars in our sample, SDSS
J142625.70+575218.4, would certainly strengthen this scenario). Since
the total amount of helium present in such stars is probably much
smaller than what is expected from ''normal'' stellar evolution, they
would have thin helium layers.

In fact, we do know a group of stars that just may have such a
characteristic: the DQ white dwarfs belonging to the ''second
sequence'' \citep{dufour05, koester06}. Indeed, while the majority of
DQ stars form a clear sequence in a $\che$ vs. $\Te$ diagram \citep[see
Fig. 12 of][]{dufour05}, several stars were found to have carbon
abundances that lie about 1 dex above the more populated ''main''
sequence \citep[this was also confirmed with an even larger sample
by][]{koester06}. The sequence where the bulk of the DQ stars are
found is naturally explained by the standard dredge-up model and it is
in agreement with the expectation that the envelopes of post-AGB stars
should have $\log q({\rm He})\equiv \log M_{\rm He}/M_{\star}$ around
-2 or -3. Only one star located on the second sequence has a mass
determination as estimated from the parallax measurement. Since it was
much more massive than all the others (1.05 \msun~vs $\langle M
\rangle$ = 0.59 \msun~for the others), it was hypothesized that maybe
stars on the second sequence were just more massive as a group and
that this could explain the higher than usual carbon
abundance. However, detailed evolutionary calculations by \citet[][see
their Figure 1]{brassard07} showed that massive white dwarf evolution
cannot account for the high abundances in the coolest DQ white dwarfs
and that thin helium envelope model could be a better
explanation. Could these special DQ white dwarfs be cooled down
versions of the hot DQs~? We believe this is a very plausible
possibility that deserves more scrutiny.

To summarize, we propose that the Hot DQs represent a short phase in
what appears to be a new channel in stellar evolution. Hence, in
addition to the general evolutionary sequence PG1159-DO-DB-DQ, there
could be another one involving H1504+65-like stars and the hot DQ
white dwarfs. This is schematically shown in Figure 
\ref{fg:f7}. Star like H1504+65, showing initially a mixed C
and O atmosphere, would show a carbon dominated atmosphere after an
intermediate phase in which it would have been observed as a DO or DB
white dwarf. With further cooling, the small amount of helium
believed to be diluted in the massive carbon envelope would rises again to
the surface and the star then would appear as a highly polluted DQ white
dwarf. Evolutionary calculations are awaited to quantitatively test
the various aspect of this scenario. 

\subsection{Other scenarios}

Are there other scenarios that could also explain the existence of
hydrogen and helium deficient objects~? One could argue that the
merging of two CO white dwarfs with a total mass less than 1.4
\msun~could produce such a thing. Or perhaps the presence of a close
companion may help to strip away the superficial layers of a white
dwarf, exposing the core. Imaginative theorists may also succeed in
finding several other ways to explain the weird surface composition of
the Hot DQ. However, there is one factor that we find very difficult
to reconcile with the above mentioned hypotheses (or any other kind of
scenario for that matter): why do they all have effective temperatures
between 18,000 and 23,000 K~? For all alternative scenarios we could
conceive of, the probability of finding a star outside this temperature
range is always greater than finding one inside. Unless future surveys
actually succeed in finding several new carbon-dominated atmosphere
white dwarfs outside this temperature range, our proposed scenario
involving episodes of diffusion and convective mixing along the
cooling sequence remains, in our view, the most likely.

The exact evolutionary history of the Hot DQs and H1504+65, related or
not, is still unknown. As was perhaps the case with H1504+65, it may
also be possible for the mass loss phases of white dwarfs of ordinary
mass reaching C-O cores to lose most or all of their helium layers,
exposing the naked core of a former AGB star. Alternatively, a
series of five papers
\citep{garciaberro94,ritossa96,garciaberro97,iben97,ritossa99}
calculated the evolution of 9, 10, 10.5 and 11 solar masses
stars. Only the last case made it to the Fe peak and would explode as a
supernova. The models in the 9-10.5 \msun~range produce massive cores
at the end of the helium-burning phase, and rise in temperature enough
to ignite carbon in a series of weak flashes, burning the carbon
core stably, and producing ONeMg cores of $> 1.2$ \msun. During the
second asymptotic giant branch phase, there are good chances that
almost all of the helium (as well as all of the hydrogen) would be
lost. This would leave a carbon-oxygen envelope/atmosphere that could
explain the origin of carbon-dominated atmospheres white dwarfs, if the
white dwarfs are massive. The following spectral and chemical
evolution of the surface could then have followed a similar path as
described above in order to be found only between 18,000 and 23,000 K.

At face value, our $\logg$ estimations generally favor a
stronger-than-usual mass loss from a C-O core star, as for only one star
(SDSS J142625.70+575218.4) is there a good case for $\logg$ around
9. For the others, $\logg$ near 8 gives the best fits, while $\logg$
as high as 8.5 can be ruled out for most. It is possible that both
mechanisms work, and that SDSS J142625.70+575218.4 indeed came from a
progenitor higher than 8 \msun~while the others have a different
origin. It is thus important to firmly establish whether stars
initially as massive as 9-10.5 \msun~can produce white dwarfs with the
properties that we observe. However, given the many uncertainties
cited above, in particular the Stark broadening of C~\textsc{ii} lines
and the UV flux level, our surface gravity determinations are only
preliminary and we will refrain from drawing any definitive
conclusions at this point.

\subsection{Pulsation Survey}

One additional interesting aspect of Hot DQ white dwarfs that has not
escaped our attention is that past models of carbon-atmosphere white
dwarfs in the range of effective temperature where the real ones are
found \citep[see][]{fontaine76} are characterized by an important
outer superficial convection zone, very similar to that found in the
pulsating DB (centered around $\Te \simeq$ 25,000 K) or in the
pulsating DA stars (found around $\Te \simeq$ 12,000
K). Hence, it follows that carbon-atmosphere white dwarfs could also
excite pulsation modes through the same partial ionization/convective
driving phenomenon that is at work in these two distinct families of
pulsating white dwarfs.

With this background in mind, \citet{fontaine08} recently investigated
the asteroseismological potential of Hot DQ's with the help of full
nonadiabatic calculations. They searched for possible instability
regions in parameter space defined by the effective temperature, the
surface gravity, the C/He ratio, and the convective efficiency. Given
the right location in parameter space, they found that some Hot DQ
stars should show pulsational instabilities against gravity modes. In
some cases, dipole ($\ell$ = 1) $g$-modes with periods in the range
100$-$700 s can be excited. The blue edge of the predicted instability
strip is hotter for higher gravity objects, while increasing the C/He
ratio in the atmosphere/envelope tends to extinguish pulsational
driving.

In parallel to this effort, an independent observational search for
luminosity variations in the current sample of Hot DQ white dwarfs was
initiated by \citet{montgomery08} at the McDonald Observatory.
In February 2008, they monitored the six accessible stars in the current
sample (SDSS J010647.92+151327.8 through SDSS J161531.71+454322.4 in
Table 1) and found overwhelming evidence for variations in the light
curve of SDSS J142625.70+575218.4 (see their Fig. 2), thus uncovering
the prototype of a new class of pulsating star. At the same type, no
variations with amplitudes above the detection threshold were observed in
the five other targets. The discovery that SDSS J142625.70+575218.4 is a
pulsating star is an extremely significant finding as it opens the door
to the application of the tools of asteroseismology to the further study
of Hot DQ white dwarfs. In essence, a new chapter in asteroseismology
has started with that discovery.

We take advantage here of our preliminary determinations of the
atmospheric properties of Hot DQ stars summarized in Table 1 to follow
up on the exciting discovery of \citet{montgomery08}. We compute,
for each of the nine objects, a dedicated full stellar model using our
estimates of $\Te$, $\logg$, and the C/He ratio. We assume a
structure with an envelope specified by a uniform chemical composition
defined by that C/He ratio, on top of a C/O core in equal proportions by
mass. For the purposes of the driving/damping process, only the chemical
composition of the envelope is of relevance \citep[see][]{fontaine08}.
We finally vary, for each target, the assumed convective efficiency in
the model from the so-called ML2 version of the mixing-length theory, to
the more efficient ML3 flavor. Hence, for each star in Table 1, we
constructed two stellar models specified by the given values of
$\Te$, $\logg$, and C/He, one with ML2 convection and the other
with ML3 convection. For those objects with no tabulated value of C/He,
we used the visibility limits quoted in Subsection 4.2; this maximizes
the possible He content and therefore the possibility of pulsational
driving.

We analyzed each equilibrium model using a full nonadiabatic approach.
We find that eight stars out of nine -- those with values of
$\logg$ = 8.0 in Table 1 -- are not expected to pulsate. This is mostly due
to the fact that their atmospheres/envelopes appear too carbon-rich to be
able to drive pulsations at their estimated values of $\Te$.
Their effective temperatures are indeed significantly higher, even
taking into account the uncertainties given in Table 1, than the
expected theoretical blue edges for a surface gravity $\logg$ = 8.0 and
the various C/He ratios involved. On the other hand, we find that a
model of SDSS J142625.70+575218.4 with $\Te$ = 19,830 K, $\logg$
= 9.0, C/He = 1, and ML3 convection is able to excite dipole $g$-modes in
the interval 163.9$-$293.8 s. While this falls somewhat short of the
period of 417.7 s found for the dominant periodicity in SDSS
J142625.70+575218.4 \citep{montgomery08}, this should not be cause
for concern at this stage because our estimations of the atmospheric
parameters of that faint pulsator need to be improved as indicated
above. We plan to actively pursue that endeavor with the MMT. For the
time being, we conclude that the expectations that we can infer from
nonadiabatic pulsation theory fall nicely in line with our
determinations of the atmospheric parameters of Hot DQ white dwarfs in
conjunction with the interesting results of \citet{montgomery08}.

\section{SUMMARY AND CONCLUSIONS}\label{conclusion}

We presented a detailed spectroscopic and photometric analysis of the 9
Hot DQ white dwarfs with carbon dominated atmospheres that have been
discovered in SDSS DR4. Strong upper limits on the abundance of hydrogen
and helium are obtained from the absence of H$\beta$ and He~\textsc{i}$
\lambda$4471, while in a few cases, small amounts of hydrogen or helium
are found in a carbon dominated atmosphere. The presence of a magnetic
field is detected in one star, while the spectra from two others suggest
the possibility of the presence of a weak field as well. Our analysis
also reveals that these stars are all situated in a narrow range of
effective temperature ($\sim$18,000-24,000 K) and that all, except
one, seem to have normal surface gravity (around $\logg$ = 8.0).
Admittedly, the precision of that result is very uncertain due in part
to the low signal-to-noise ratio of the data. Additionally, this first
generation of carbon atmosphere models must be considered preliminary,
as many factors might influence our results (opacity sources in the UV,
broadening theory etc.). We retain two main hypothesis for the origin of
these stars i) They might be the results of C-O core white dwarfs that
have experienced an unusually high mass loss at the end of the AGB phase
or ii) be the progeny of massive stars that had carbon burning and
produced O-Mg-Ne cores white dwarfs. If the latter is the case, the hot DQ
stars may ultimately help to better define the limiting mass at which
massive stars explode as a type II supernova as well as the birth rate
of neutron stars.

Future work on these objects will be dedicated to the analysis of higher
signal-to-noise ratio observations. Follow up spectroscopic observations
with the Mount Hopkins 6.5m MMT telescope and the Mount Fowlkes Hobby
Eberly Telescope are underway and will be present in due time. Our next
generation of model atmospheres for carbon dominated compositions should
include the latest Stark broadening theory of C~\textsc{ii} lines as
well as any updates on sources of opacity that are judged necessary. We
also hope to obtain time for near-UV and far-UV observations with the
Hubble Space Telescope (HST) using the Cosmic Origin Spectrograph (COS),
after the scheduled repair mission, to assess the quality of our
modeling of the UV part of the spectrum. Finally, now that we know how
to easily recognize these stars, we are paying special attention to SDSS
data releases so that the number of objects in this new interesting
spectral class is expected to increase in a near future.

\acknowledgements

We wish to thank D. Eisenstein for very useful discussions concerning
the SDSS data and the DR4 white dwarfs catalog and P. Smith for help
with the spectropolarimetric observations.  P.D. acknowledges the
financial suport of NSERC. This work was also supported by the NSF
through grant AST 03-07321. Support from the Reardon Foundation is
also gratefully acknowledged.

\clearpage

 \clearpage
 \begin{deluxetable}{lccrccccc}
 \tabletypesize{\footnotesize}
 \tablecolumns{10}
 \tablewidth{0pt}
 \tablecaption{Atmospheric Parameters of Hot DQ Stars from the SDSS}
 \tablehead{
 \colhead{Name} &
 \colhead{plate} &
 \colhead{MJD} &
 \colhead{Fiber} &
 \colhead{$T_{\rm eff}$(K)} &
 \colhead{log $g$} &
 \colhead{log C/H} &
 \colhead{log C/He} &
 \colhead{$D$ (pc)}}
 \startdata
SDSS J000555.90$-$100213.3 &  650&52143&   37& 19420 (920)&8.0& - & - & 255\\
SDSS J010647.92$+$151327.8 &  422&51878&  422& 23430 (1680)&8.0& - & 1.00& 492\\
SDSS J023637.42$-$073429.5 &  455&51909&  403& 21110 (430)&8.0& - & - & 738\\
SDSS J115305.54$+$005646.2 &  284&51943&  533& 21650 (320)&8.0& - & - & 515\\
SDSS J133710.19$+$002643.7 &  299&51671&  305& 21670 (245)&8.0& 1.67& - & 461\\
SDSS J142625.70$+$575218.4 &  789&52342&  197& 19830 (750)&9.0& - & 0.0& 245\\
SDSS J161531.71$+$454322.4 &  814&52443&  577& 20940 (460)&8.0& 1.74& - & 719\\
SDSS J220029.09$-$074121.5 &  717&52468&  462& 21240 (180)&8.0& - & - & 283\\
SDSS J234843.30$-$094245.2 &  648&52559&  585& 21550 (340)&8.0& - & - & 544\\
 \enddata
 \end{deluxetable}
 \clearpage

\clearpage

\figcaption[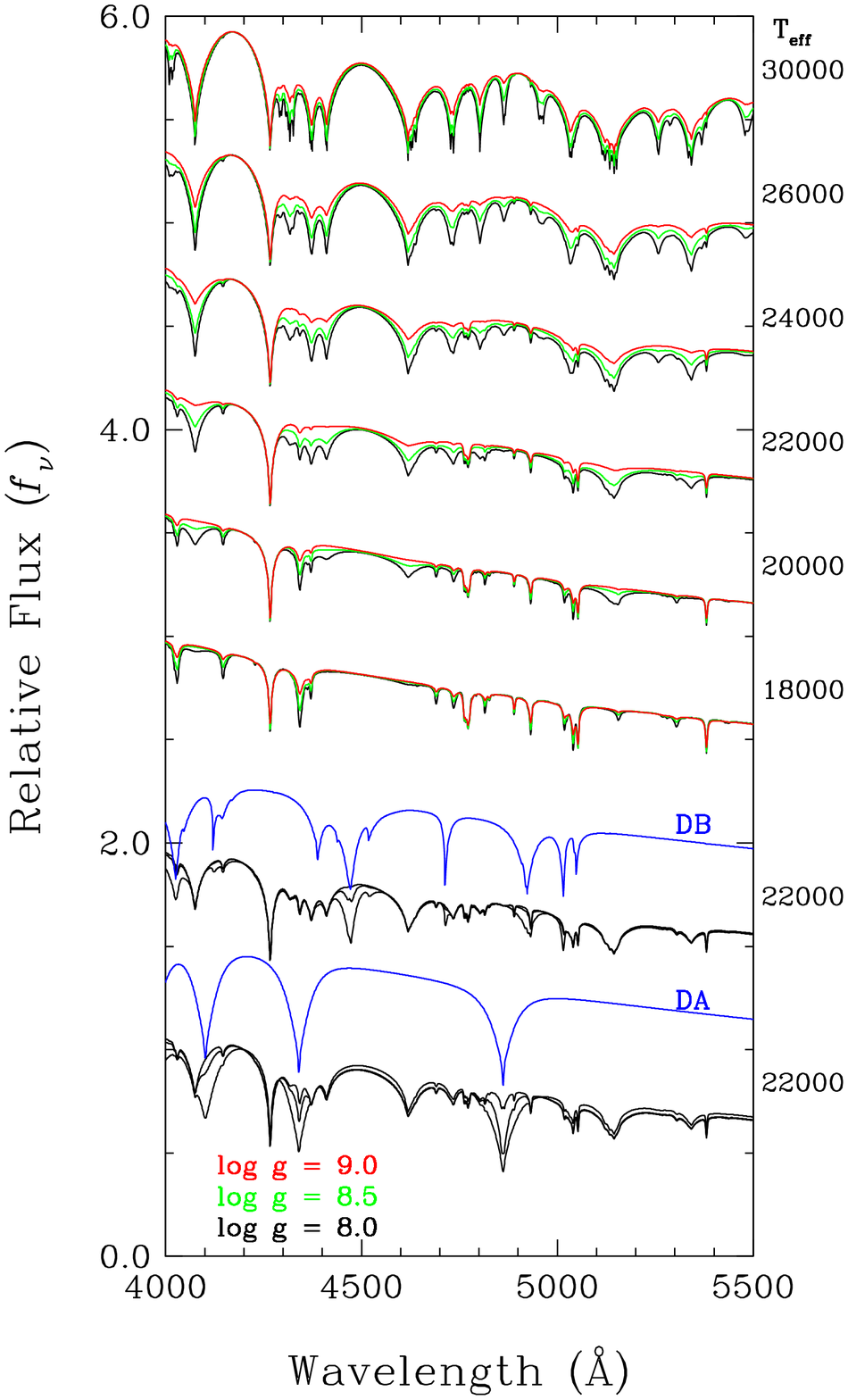] {Representative synthetic spectra of Hot DQ white
dwarfs taken from our model grid for various effective temperatures,
gravities, helium and hydrogen abundances. The spectra are normalized to
unity at 4200 \AA\ and offset by an arbitrary factor for clarity. The
labels indicate the effective temperature, while the color of each
spectrum corresponds to a different gravity, as indicated in the lower
left corner. The bottom two examples show the influence of trace
helium ($\che$ = 2, 1 and 0) and trace hydrogen ($\ch$ = 2, 1 and
0) while in blue, we show the log g = 8 synthetic spectra for pure
helium atmosphere (DB) and pure hydrogen atmosphere (DA).
\label{fg:f1}}

\figcaption[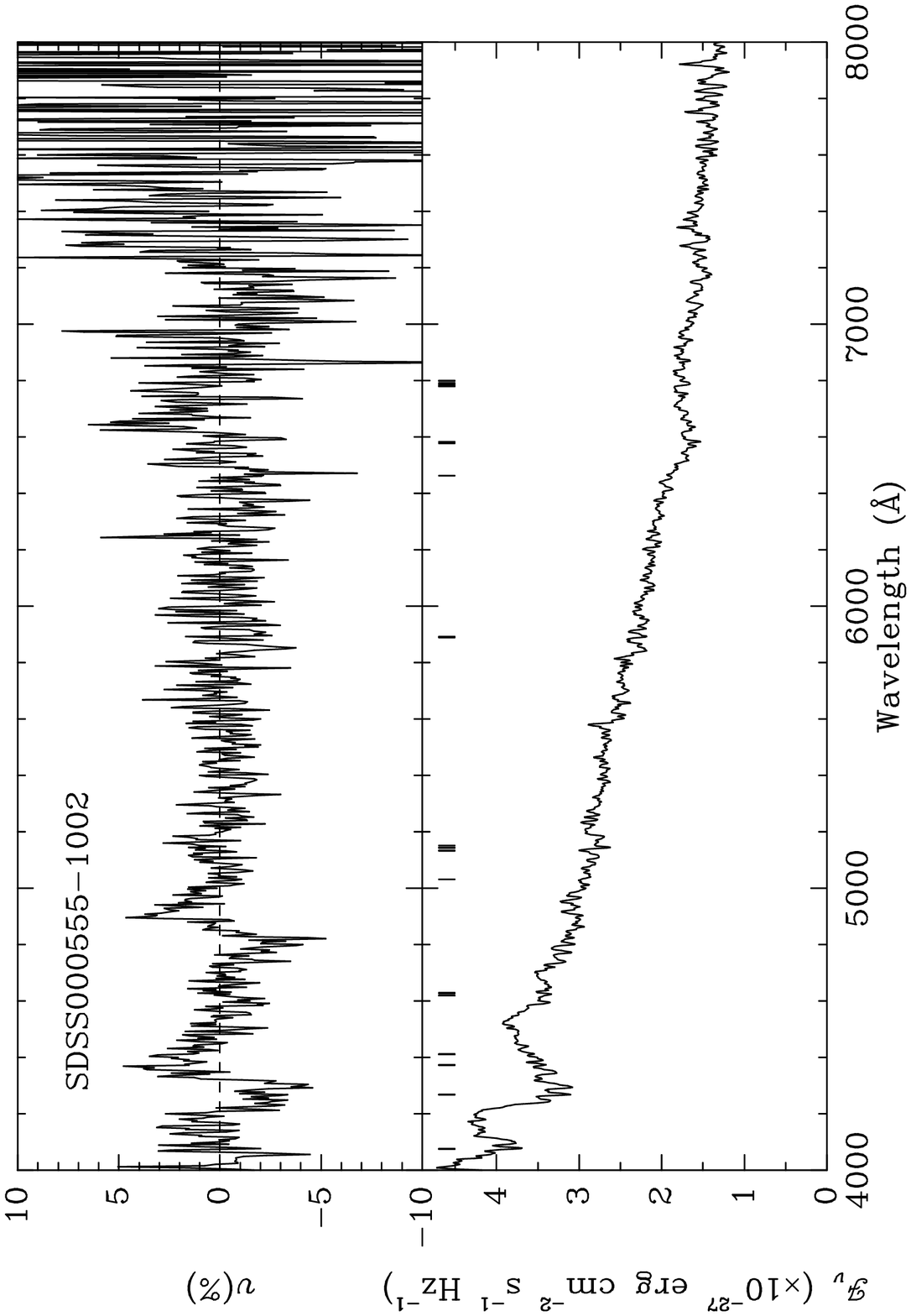] {Observed circular polarization (top panel) and 
the corresponding spectrum obtained with a spectral resolution 
($\Delta\lambda\sim17$~\AA). The thick marks indicate the position of the 
strongest C~\textsc{ii} features.
\label{fg:f2}}

\figcaption[f3.eps] {Fits to the energy distribution (right panel) and
carbon lines (left panel) for all carbon dominated DQ white dwarfs in
our SDSS sample. The $ugriz$ photometric observations are represented
by error bars, while the average model fluxes are shown by filled
circles. The derived atmospheric parameters are indicated in each
panel. We have applied for clarity a three-point average window
smoothing in the display of the spectroscopic data.
\label{fg:f3}}

\figcaption[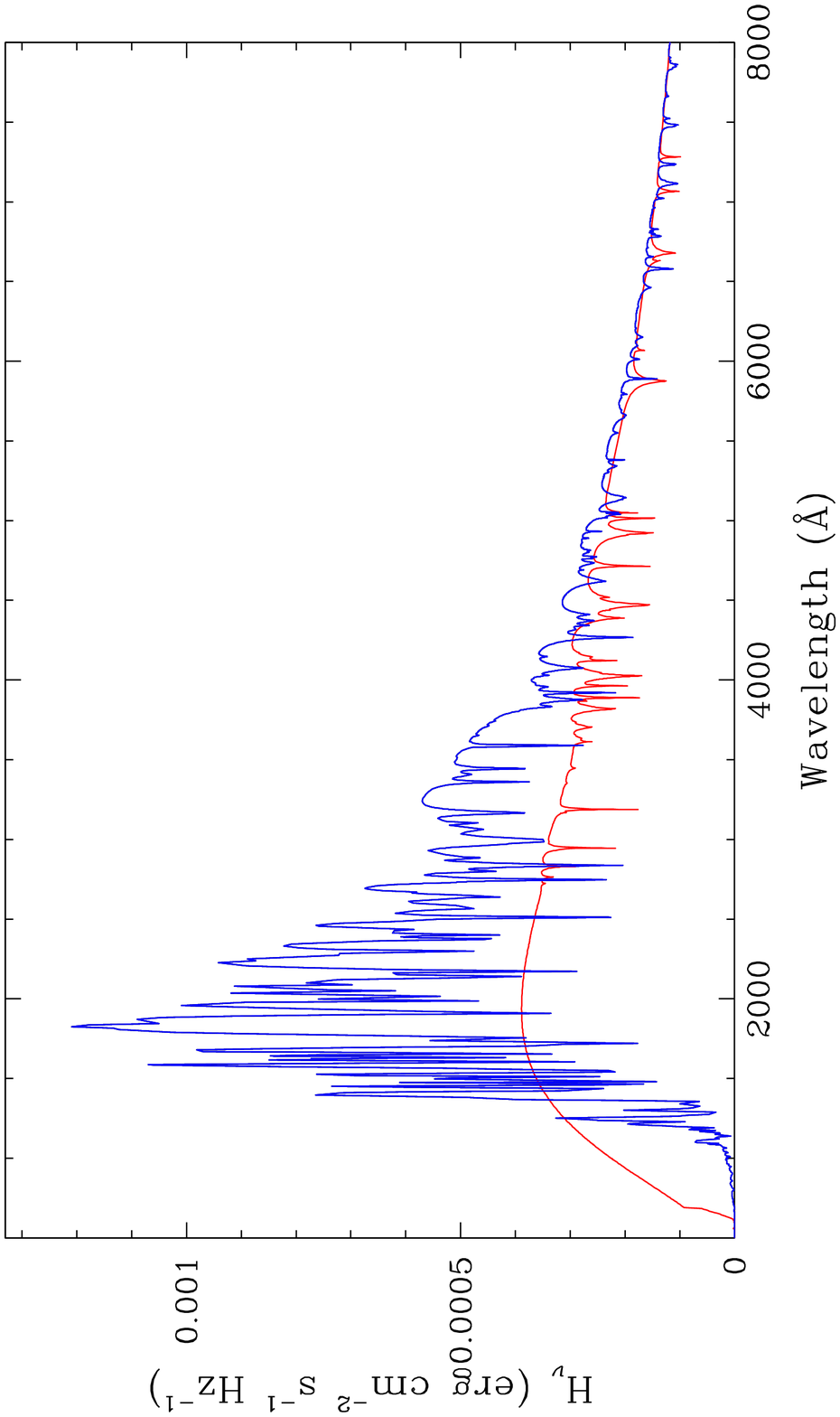] {Comparison of a pure helium DB model (red line)
with a pure carbon model (blue line) for $\logg$ = 8.0 and $\Te$ =
22,000 K. Note that $\int H_{\nu} d\nu$ is the same for both models (=
$\sigma \Te ^4/4\pi $).
\label{fg:f4}}

\figcaption[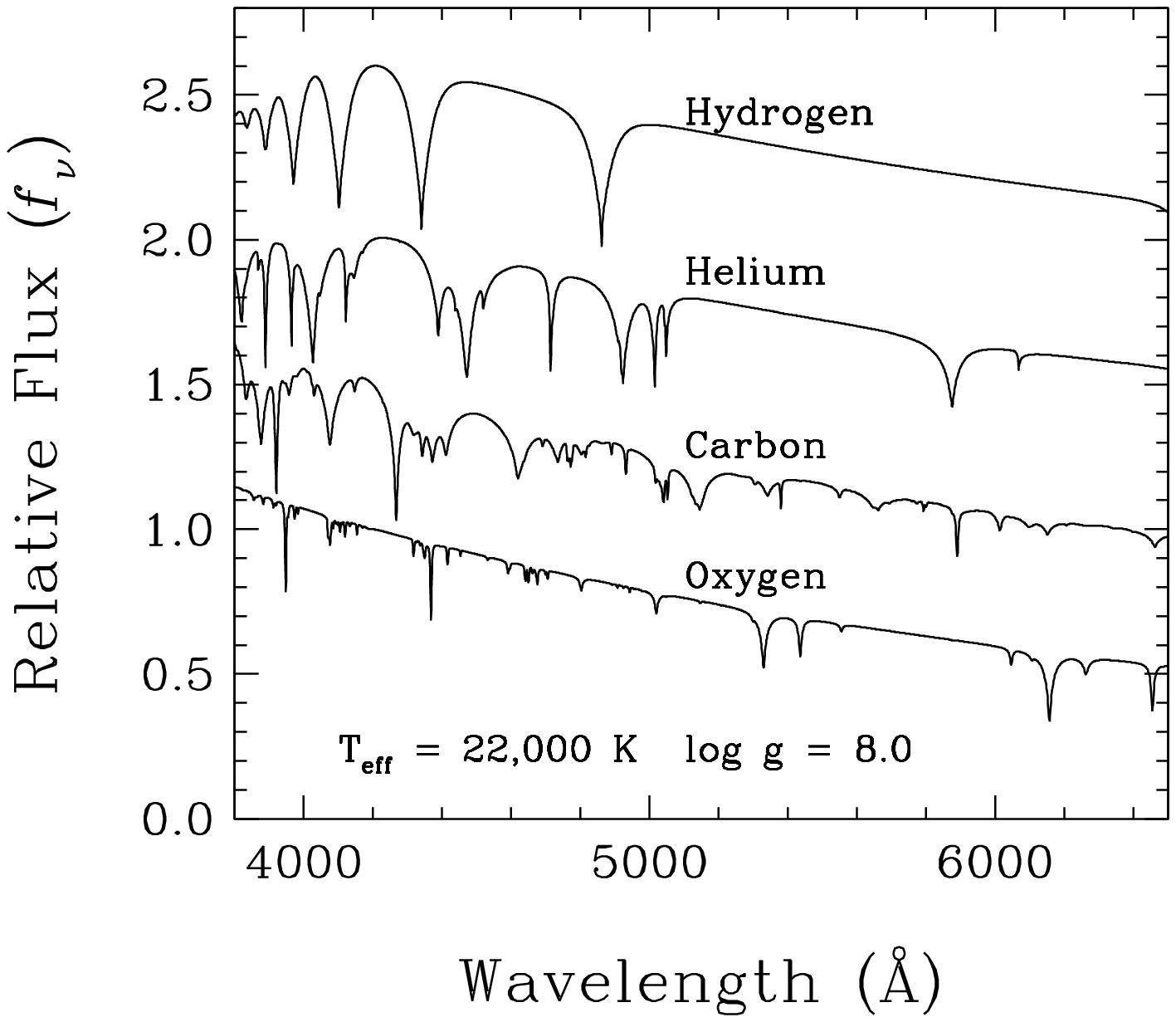] {Comparison of the optical spectra of pure
composition atmospheres composed of hydrogen, helium, carbon and oxygen
for $\logg$ = 8.0 and $\Te$ = 22,000 K.
\label{fg:f5}}

\figcaption[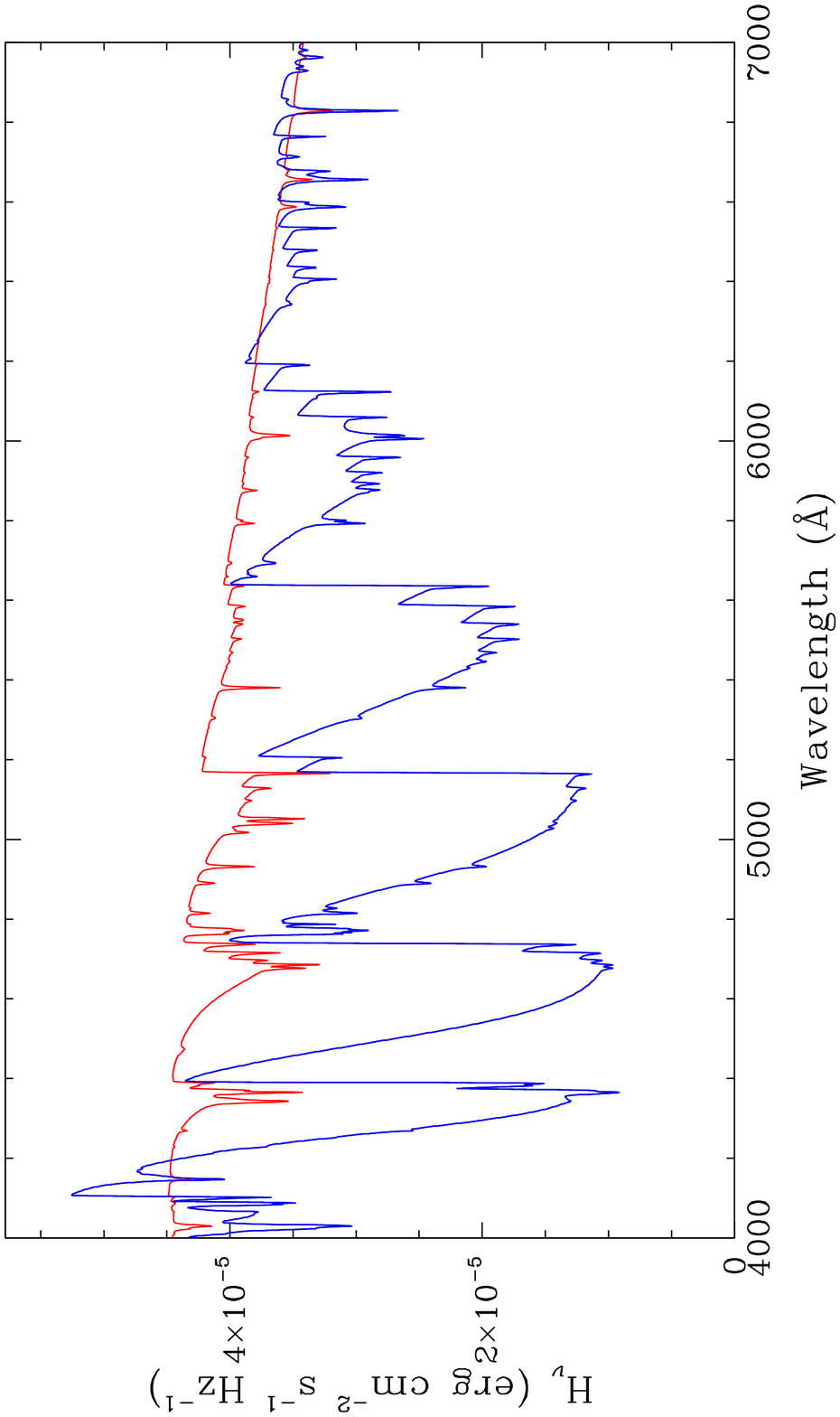] {Comparison of synthetic spectra of a pure carbon
atmosphere models (blue) with a $\che$ = -3.0 model (red) for $\Te$ =
10,000 K and $\logg = 8.0$. Spectra with extremely strong carbon bands 
similar to that predicted in the pure carbon model should have been
easily noticed in SDSS if they existed.
\label{fg:f6}}

\figcaption[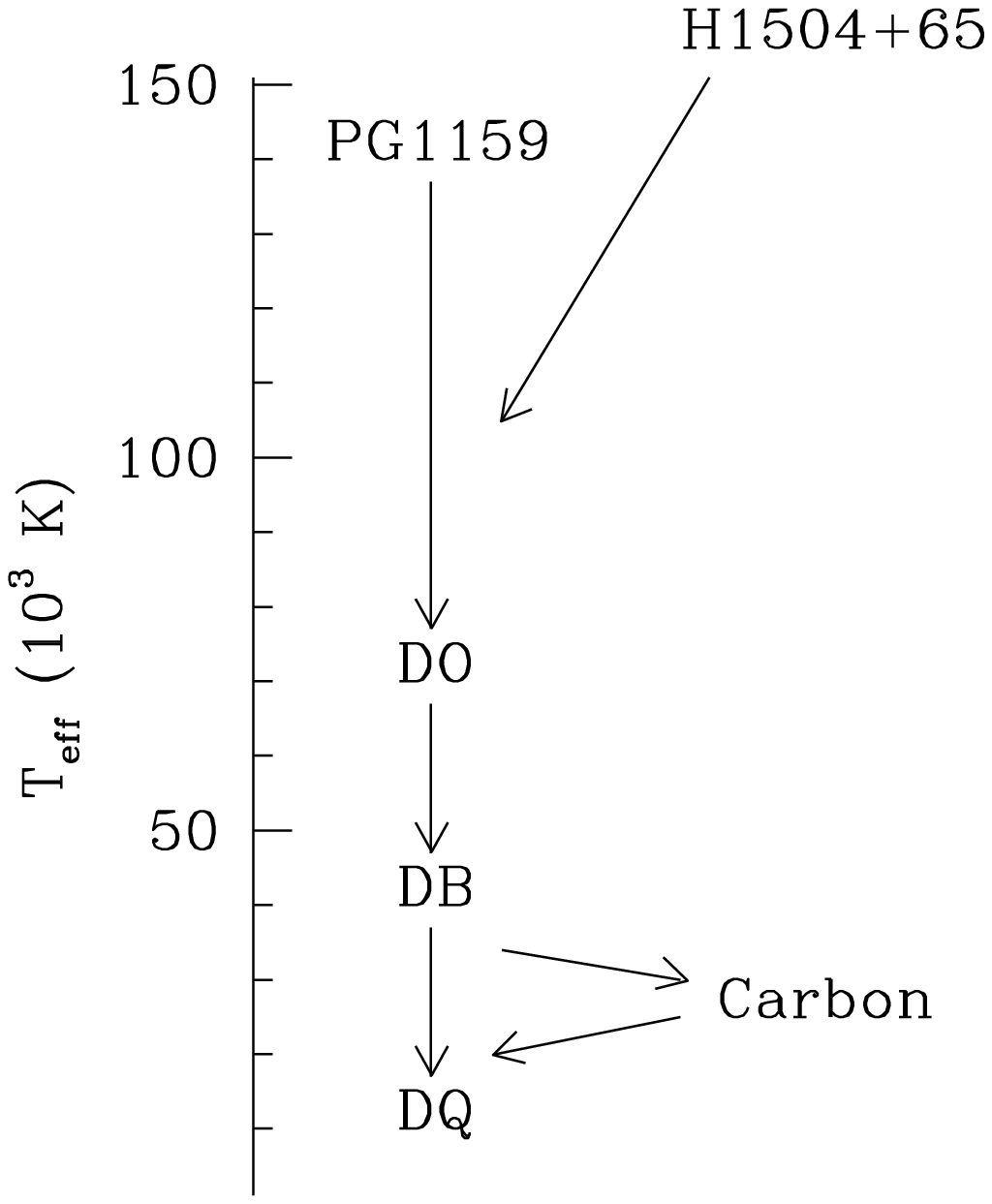] {Schematic representation of our proposed evolutionary
scenario to explain the existence of carbon dominated atmosphere white dwarfs.
\label{fg:f7}}

\clearpage
\begin{figure}[p]
\plotone{f1.eps}
\begin{flushright}
Figure \ref{fg:f1}
\end{flushright}
\end{figure}

\clearpage
\begin{figure}[p]
\plotone{f2.eps}
\begin{flushright}
Figure \ref{fg:f2}
\end{flushright}
\end{figure}

\clearpage
\begin{figure}[p]
\plotone{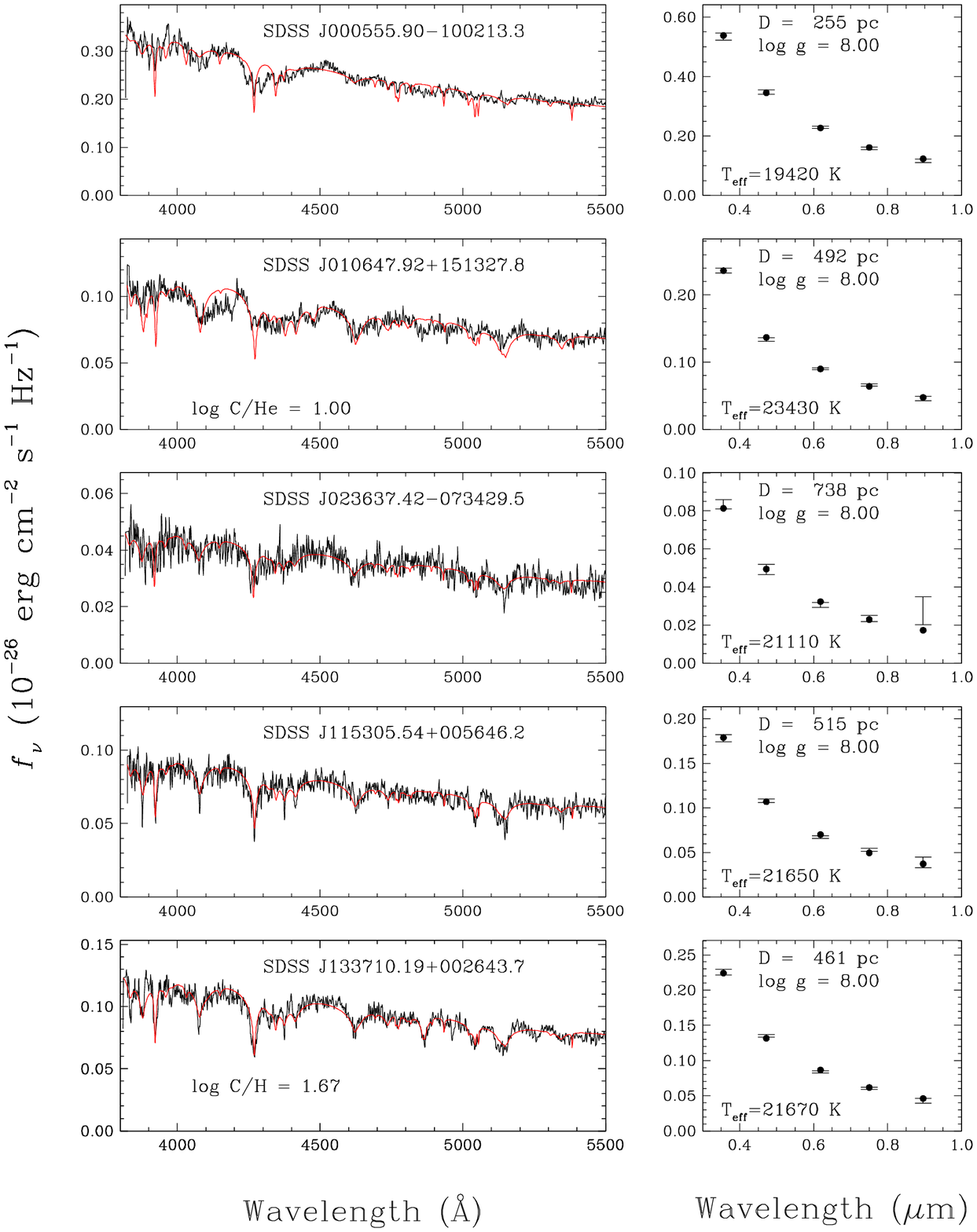}
\begin{flushright}
Figure \ref{fg:f3}
\end{flushright}
\end{figure}

\clearpage
\begin{figure}[p]
\plotone{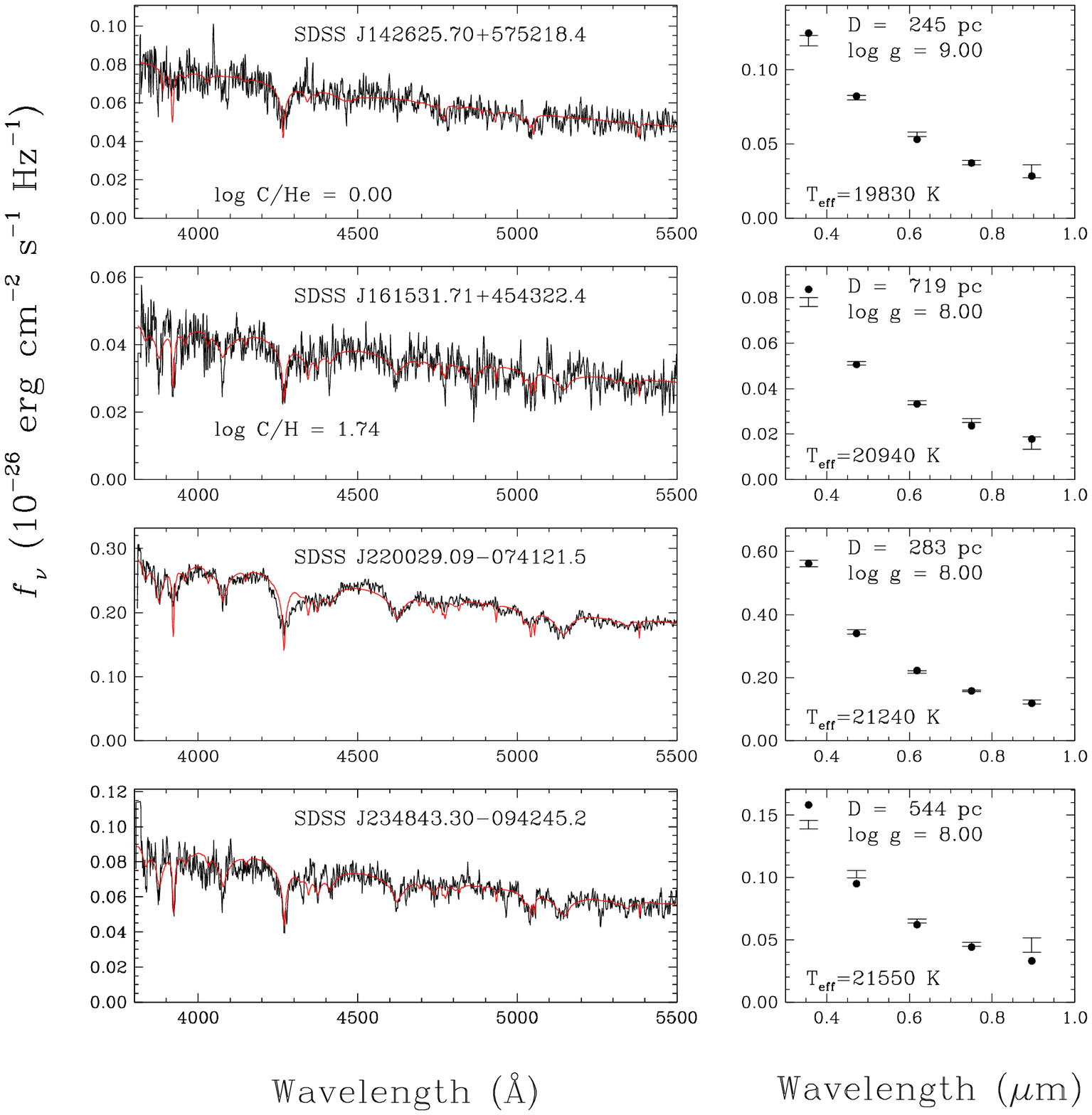}
\begin{flushright}
Figure \ref{fg:f3}
\end{flushright}
\end{figure}

\clearpage
\begin{figure}[p]
\rotate{90}
\plotone{f4.eps}
\begin{flushright}
Figure \ref{fg:f4}
\end{flushright}
\end{figure}

\clearpage
\begin{figure}[p]
\plotone{f5.eps}
\begin{flushright}
Figure \ref{fg:f5}
\end{flushright}
\end{figure}

\clearpage
\begin{figure}[p]
\plotone{f6.eps}
\begin{flushright}
Figure \ref{fg:f6}
\end{flushright}
\end{figure}

\clearpage
\begin{figure}[p]
\plotone{f7.eps}
\begin{flushright}
Figure \ref{fg:f7}
\end{flushright}
\end{figure}

\end{document}